# You Are Okay:
## *Towards User Interfaces for Improving Well-being*

Pedro F. Campos
*Madeira-ITI, University of Madeira, Campus da Penteada, Funchal, Portugal*
*pedro.campos@m-iti.org*

Abstract: Well-being is a relatively broad concept which can be succinctly described as the state of being happy, healthy or successful. Interesting things happen when bridging user interface design with the psychology of human well-being. This position paper aims at providing a short on reflection the challenges and opportunities in this context and presents concrete examples on how to tackle these challenges and exploit the existing design opportunities.

## 1 INTRODUCTION

Well-being is a relatively broad concept which can be succinctly described as the state of being happy, healthy or successful. Being happy (or feeling successful) is mostly related to mental well-being, while being healthy is typically associated to the physical or physiological well-being. However, this correspondence is not fully straightforward. It is the complete interplay between these three qualities that effectively brings us to a state of full well-being. Interaction design, on the other hand, is the practice of designing interactive digital products, systems, and services focuses on creating engaging experiences for a wide variety of audiences, contexts and domains. The design of any sort of interactive experience is an exercise that does not lend itself to any dogmas or prescriptive solutions.

Skeptics are correct when they claim that the history of mankind is a story of missed opportunities. The same could be said about the act of designing user interfaces. Any user interface is something that is constantly under construction, but also, in parallel, in a state of constant destruction. It is often said that the best designers are those who force themselves into inventing, evaluating and evolving the largest possible number of design ideas. In this context, interesting things happen when bridging user interface design and the psychology of human well-being. This position paper aims at providing a short on reflection the challenges and opportunities in this context and presents concrete examples on how to tackle these challenges and exploit the opportunities.

A serious amount of research has been dedicated to improving mental well-being through mobile technology and interactive experiences. Augmented reality, for instance, has become quite popular in contexts such as the classroom (Campos and Pessanha, 2011). Lathia et al. (2013) defend that it is possible to make inferences about use contexts, physical activities, and mental states (including emotions and stress) using data from smartphone sensors (Lathia et al., 2013). These are also privileged devices to provide tailored advice, support goal setting, help users plan and chart their progress, and send personalized emails or SMS reminders. Their system (called EmotionSense) automatically recognizes who is speaking and what the speaker is feeling, using classifiers that run locally on phones. They had good results in predicting emotions from speech and believe this solution obviates the need for self-input. It still needs work and a perfection of technology, however, to be fully reliable. They also test another application called SocioSense, that applies a gamification method to the social communications of the user (Lathia et al., 2013).

Social networks, essentially Facebook, have also been studied from the perspective of emotion-sharing. Buechel and Berger (2012) set out to accomplish two goals: to understand if "emotionally unstable" people post personal stories on Facebook more than others, and to know what types of sharing possibilities have a bigger positive impact on the user's well-being. Essential findings showed that users were classified as Emotionally Stable and Unstable based on a well-documented survey/scale called Big Five Personality Inventory (John and Srivastava, 1999).

Journaling is a means of emotion expression. It prevents the inhibition of emotion which is harmful as a chronic stressor (Ullrich and Lutgendorf, 2002).

Expressing Trauma-Related emotions in a safe environment enhances feelings of control and mastery over traumatic events (Ullrich and Lutgendorf, 2002).

Journaling also allows for cognitive processing of critical life situations where subjective meanings of the world have been questioned. An experiment compared the effects of writing about emotions, facts or both and realized that the combination of both presents the best result. Researchers tried to compare emotional disclosure (journaling) versus cognitive disclosure and both. They also compared differences in the content of the writing and their outcomes. The group that wrote about both emotions and facts increasingly wrote about cognitive processing topics (understanding the problems, rather than stressing on the emotions) (Ullrich and Lutgendorf, 2002).

Ma and colleagues (2012) propose a novel framework called MoodMiner for assessing and analyzing mood in daily life. MoodMiner uses mobile phone data – mobile phone sensor data and communication data (including acceleration, light, ambient sound, location, call log, etc.) - to extract human behavior pattern and assess daily mood. Experimental results on 15 users for a month showed an effective assessment of the system to decipher daily moods objectively with minimal user intervention. The sensor types used to model the system application include: accelerometer - used to detect the user's physical environment; sound sensor - used to sense background sound and the user's voice that can both discern the mood state of the user; location - used to provide context information for user behavior. The application was designed to have a set of daily behavior features extracted from mobile sensor data and communication data.

Because of the subjective nature of mood, sensor data, communication data and the failure to reflect mood swings through historical mood data - the analyzed mood may be incorrect. Further, users show a significant difference in daily behavior style and phone usage pattern. The model discussed by Ma et al. has the potential to be successful, however it would involve the dedicate use of mobile phones to keep it from failing (Ma et al., 2012).

To achieve better levels of human well-being, both physical and mental well-being, one can use a design approach targeted at positive behavior change. Behavior change is a central objective in public health interventions, with an increased focus on prevention prior to onset of disease. Due to many factors, such as the increasing popularity of activity trackers, interest has been rising on approaches where change happens through reflection. However, this approach has one major drawback: there is an excessive reliance on user's motivation to explore the data. Faced with this problem, interaction designers have been investigating strategies for behavior change that do not require thoughtful effort to adopt more positive behaviors.

Nudges have been revealing as the most promising approach; they endeavor to make judgments and choices easier through incentives, reinforcement and unforced suggestion right when decisions take place behavior change. A more radical approach to nudges has been the so-called "pleasurable troublemakers" (Laschke et al., 2014). Keymoment illustrates this approach. It is a gadget that holds one's car and bicycle keys. Upon departure, the user makes a choice about his mode of transportation. If he chooses the car, Keymoment drops his bicycle keys on the floor, nudging him to reconsider his choice.

## 2 OUR APPROACH

Our approach to creating novel interactive experiences for improving well-being is based on the concept of intervention. In psychology, interventions are actions performed to bring about change to people, and modern applied psychology is very rich in intervention strategies and techniques. A positive psychological intervention is an exercise, or set of exercises, which have been shown in lab experiments to increase positive emotions, or other desirable states.

Therefore, we typically begin the process by identifying the type of problem we are dealing with, and then we conceptualize a series of possible interventions in a technology-independent way. This requires, naturally, the deep involvement of psychologists in the team, who play an active role in the design of the interactive experiences, from day one until the successive roll-outs and in loco evaluations of such experiences.

Through our own experience during the last four to five years, we have tailored the typical HCI design process to include a few steps that seem to help designers to open way for improved human well-being. There are three "tasks" that are always conducted:

**1. Watch a Movie.** This implies getting popcorns and drinks, forcing the entire team to spend time together in front of a distinct "reality", with the specific goal – pre-established! – of using the movie's characters as actors in the product brainstorming.

**2. Make a Field Trip.** When designing user interfaces for all things related to well-being, nothing replaces the value of a real, focused, honest field trip. It is especially important to carefully document all aspects of the field trip, as it is equally critical to maintain that documentation in well-printed large posters scattered throughout the work space of the team.

**3. Design a Cartoon.** This technique is similar to scenarios or storyboards, but requires an explicit creation of a cartoon. Some of our team members have recently started to use tools such as ComicLife and "Instagram-like" apps like Prisma to streamline this process and make the results more appealing and inspiring. Some work analysis methods also employ similar techniques to this one (Campos et al., 2013).

Providing examples of the type of interventions and interactive experiences we have developed is probably the best way to illustrate this approach.

## 3 EXAMPLES

### 3.1 Promoting Well-being by Social Media-triggered Routine Breaks

With the rise of social networks, connectivity and media consumption have seen dramatic changes. In this work, we focus on understanding how mental well-being relates to people's routines, what dangers and recommendations we should be aware of, and what opportunities exist to leverage current technologies in order to improve people's lives.

As a contribution to HCI based on our research insights, we designed a solution that promotes daily mental well-being. Spark is a mobile application where everyone in the world participates in a single challenge every 25 hours. Each challenge asks the user to perform an activity, capturing that moment with a photo, and then sharing it.

As the user participates in different challenges over time, a record of his activities is always accessible. This encourages the user to relive positive past experiences, and take note of his accomplishments. This is an example of how mobile technology can be deployed with the specific goal of improving people's mental well-being. The results seem promising as this research project progresses.

### 3.2 Mitigating Excessive Consumerism

Excessively consuming essentially futile or unnecessary assets has turned into a true epidemic, at least in the richest and more developed countries in the world. These consuming habits, when excessive, can really lead to serious problems, of psychological nature, therefore converting excessive consumption into an effective public (mental) health problem. We studied the factors that drive people to consume more or less, and the psychologists in our team conducted an empirical study in order to inform the design of novel interactive technologies for mitigating this problem. In this context, we proposed #LookWhatIDidNotBuy as a new psychological counseling app that promotes the social media sharing of the photos of goods that the user managed not to buy, thus resisting the temptation. The app also provides advice using positive reinforcement, daily challenges, and tips.

The relevant aspect of this approach stems from the fact that our design goes against the dominant narrative of goal-setting apps and goal-setting theory, advocating that sharing the media of goods not bought can induce positive behavior change.

### 3.3 Exploiting Hypnagogic States using VR to Improve Creativity

Creativity is actually a very easy way to achieve well-being. Visual art represents a powerful resource for mental and physical well-being. Studies have convincingly demonstrated that the production of visual art improves effective interaction between parts of the brain. Karen Robinson, for instance, is a very interesting artist who conducts post-traumatic growth. She uses art, creative writing and photography as the main tools for therapy. Well-being is not just about positive affective states. It often involves cognitive evaluations of the conditions of one's life (e.g., overall life satisfaction). One of the problems faced by today's generation of knowledge workers is the lack of creativity, experienced by writers in the form of writer's block. The perception of becoming unsuccessful in the professional sides of our lives can lead to a decrease in our well-being - this reflects both mentally and physically. In this project, we explore Virtual Reality as a tool to kickstart hypnagogic states. In the borderlands between wakefulness and rest there is a strange and fascinating state of consciousness characterized by dream-like visions and strange sensory occurrences. Psychologists call this stage "hypnagogia," but centuries before they created a term for it, artists were already using the hypnagogic state to tap into some of their best ideas.

### 3.4 Supporting Vulnerable Populations

Interaction design for well-being can also shape itself as a social movement with impact over more vulnerable segments of the population. In a 2015 project, our team embarked on a quest to improve the mental well-being of underserved youths in a local school. Minority groups are the fastest growing demographic in the U.S. In addition, the poverty level in the U.S. is the highest it has been in the last 50 years. The community, therefore, needs more research addressing this user segment. We studied how underserved youths react when presented with different UI designs aimed at promoting creative writing (Gonçalves et al., 2015). The act of creative writing per se can become the driver of change among underserved teenagers, and researchers should strive to discover novel UI designs that can effectively increase this target group's productivity, creativity and mental well-being. Using MS Word as baseline, we analyzed the influence of a Zen-like tool (a tool designed at our lab, and called PlaceToWrite), a nostalgic but realistic typewriting tool (Hanx Writer), and a stress-based tool that eliminates writer's block by providing consequences for procrastination (Write or Die).

Our results suggest that the Zen characteristics of our tool PlaceToWrite were capable of conveying a sense of calm and concentration to the users, making them feel better and also write more. The nostalgic Hanx typewriter also fared very well with regard to mental well-being and productivity.

There have been other interventions designed, developed and evaluated by our team, including: Improving family life; improving the life and health of the aging population; dealing with work, multitasking and distractions; and primary prevention of violence in teenage dating.

## 4 CONCLUSIONS AND OPEN QUESTIONS

As with all scientific problems, many questions remain permanently open to improvements. How to improve the levels of collaboration between teams of psychologists, interaction designers, marketing experts and engineers? How can we better evaluate the quality of the interactive experience, the same way psychologists assess the quality of their interventions? Also, how can interaction designers target experiences that are specifically tailored to each user's problems and needs?

As interaction designers, we have the power to use technology as a vehicle for transformative experiences. Information technologies have deeply transformed our world, and are now weaved into many aspects of our daily life. Simply recognizing the potential of interaction design for improving human well-being is not enough. Designers need principles, guidelines, successful case studies from which they can obtain inspiration leading to an ever-increasing number of interactive experiences that can be effective in attaining the eternal goal of an improved well-being.